\documentclass[11pt]{article}
\usepackage{jheppub}
\usepackage{pstool}
\pdfoutput=1
\usepackage{color}
\usepackage{float}
\usepackage{array}
\usepackage{amssymb}
\usepackage{amsmath}
\usepackage{amsthm}
\usepackage{graphicx}
\usepackage{caption}
\usepackage[labelsep=quad]{subcaption}
\usepackage{epstopdf}
\usepackage{placeins}
	
\usepackage{epsfig}
\def\op{{\cal O}}

\def\pb[#1,#2]{\{#1, #2\}}
\def\deb[#1,#2]{[#1,#2]_{\text{D.B.}}}

\def\Or[#1]{{\text{O}}\left({#1}\right)}
\def\dotl[#1,#2]{\left\langle #1,\, #2 \right\rangle}
\def\dotlb[#1,#2]{\left\langle #1,\, #2 \right\rangle}
\def\dotlm[#1,#2]{\left[ #1,\, #2 \right]}
\def\dotp[#1,#2]{(\vect{#1} \cdot\vect{#2})}
\def\aff[#1,#2]{\hat{#1}(#2)}

\def\n4sym{{\cal N}=4 SYM}
\def\>{\rangle}
\def\<{\langle}
\def\weight[#1,#2,#3]{\{(#1),#2,#3\}}
\def\ads[#1]{$\text{AdS}_{#1}$}

\newcommand{\be}{\begin{equation}}
\newcommand{\ee}{\end{equation}}
\newcommand{\ba}{\begin{align}}
\newcommand{\ea}{\end{align}}
\newcommand{\bs}{\begin{split}}
	\def\sess\end{split}
\newcommand{\vect}[1]{{\boldsymbol{#1}}}

\def \bea {\begin{eqnarray}}
\def \eea {\end{eqnarray}}
\def \bea* {\begin{eqnarray*}}
	\def \eea* {\end{eqnarray*}}

\def \be {\begin{equation}}
\def \ee {\end{equation}}
\def \bes {\begin{equation*}}
\def \ees {\end{equation*}}

\title{Behind-the-horizon excitations from a single 2d CFT}
\author[a]{Souvik Banerjee}
\emailAdd{souvik.banerjee@uni-wuerzburg.de}
\affiliation[a]{Institut für Theoretische Physik und Astrophysik, Julius-Maximilians-Universität Würzburg,\\ Am Hubland, 97074 Würzburg, Germany} 
\author[b]{and Gideon Vos}
\emailAdd{vos@fzu.cz}
\affiliation[b]{Central European Institute for Cosmology, \\ FZU, Na Slovance 1999/2, 182 21 Prague 8, Czech Republic\\}

\date{}

\abstract{In this work, we consider the atypical non-equilibrium state found in \href{https://arxiv.org/abs/1708.06328}{[1708.06328]} which holographically represents a behind-the-horizon excitation in a black hole spacetime. The special feature of this state is that it looks like an equilibrium state when probed by a class of low-energy operators. First, we retrieve this property using the uniformization mapping in the limit of a large central charge, in the process we are able to derive rather than presume approximate thermal physics.
Furthermore, in the large-$c$ and high-energy limit, we realize these excitations as elements of the commutant algebra of a GNS-representation of the light operator algebra. Instead of analytically continuing a mixed heavy-light Euclidean correlator to a Lorentzian correlator, we identify the Euclidean correlator as a GNS-linear form and interpret the Lorentzian correlator as a vacuum expectation value of representatives of the light operator algebra on the GNS-vacuum.}
\begin{document}
	\maketitle

\tableofcontents

\section{Introduction}
The Rosetta stone towards a quality understanding of quantum gravity is to gain a proper understanding of the quantum aspects of black holes. Towards this end, holographic conformal field theories \cite{Maldacena:1998im, Witten:1998qj, Gubser:1998bc}, the special class of conformal field theories which possess a dual gravitational description at large $N$ \cite{Heemskerk:2009pn, El-Showk:2011yvt}, provide a powerful tool set. While constructing sound quantitative questions regarding black holes in full quantum gravity is at best nebulous, quantitative questions in dual CFT computations are in many cases well-defined, even if potentially arbitrarily complex. Even better, provided we ask the right questions even in a lower-dimensional controllable CFT, the general lessons learned are often expected to at least qualitatively carry over to more complex CFTs. Even if the actual computation in the most general setting still remains technically challenging, we know, in principle, where to look if the future endows us with greater computational power. This is part of a general philosophy towards the development of mathematically robust, physically relevant toy examples.

The horizon of a big black hole in AdS spacetime is understood as a consequence of entanglement between two sets of degrees of freedom of the boundary CFT \cite{VanRaamsdonk:2010pw}. While this is well visualized in the context of a two-sided eternal geometry with two entangled CFTs living on its two boundaries \cite{Maldacena:2013xja}, this entanglement is not geometrically manifest in a more realistic collapsing geometry which lacks the second boundary. Furthermore, it was independently argued using the properties of entanglement of modes near the black hole horizon that an interior of a big black hole might not actually exist \cite{Mathur:2009hf}, particularly, in a way consistent with the understanding of AdS/CFT \cite{Almheiri:2012rt, Almheiri:2013hfa, Marolf:2013dba}. 
This issue was successfully addressed in a series of seminal works \cite{Papadodimas:2012aq,Papadodimas:2013jku,Papadodimas:2013wnh, Papadodimas:2015jra, Papadodimas:2015xma} where a ``state-dependent'' reconstruction of the black hole interior was proposed and subsequently developed using an algebraic modular theory, suitably adopted for a system with a large number of degrees of freedom, namely, a large $N$ holographic field theory. In short, this construction exploits the large $N$ hierarchy in holographic CFTs which in turn results in a complexity class of the CFT operators. Based on this, the construction deals with an ``approximate'' algebra of ``simple operators'' which, given a particular pure state gives rise to a small Hilbert space. The operators in the interior are realized as the mirror operators which are commutants acting on the small Hilbert space itself. Here `small' refers to operators that generate $\mathcal{O}(1)$ amount of superfield excitations in the bulk, e.g. small products of single-trace operators in $\mathcal{N}=4$ SYM. It was found that the dimensionality of the CFT Hilbert is rich enough that for any local observable one can construct a state-dependent mirror operator that commutes with the original set of local observables. The most important point here is that both the small algebra and its commutant act on the small Hilbert space but the commutation between them in general does not hold in the full theory, namely, inside a large correlator with the number of operators scaling with $N$. 
This establishes, mathematically, the concept of black hole complementarity \cite{THOOFT1990138, Susskind:1993if} in the sense that locality is only an emergent concept relevant at low energies, i.e. inside the low point correlators. This is the main essence of the state-dependent construction. By virtue of being explicitly state-dependent, this construction evades the arguments against the interior reconstruction mentioned above.
Apart from the initial works already mentioned above, for further details on the state-dependent constructions and many subsequent developments in this direction, we refer to the review \cite{Raju:2020smc}. 

\begin{figure}[h]
	\centering
		\includegraphics[scale=0.5]{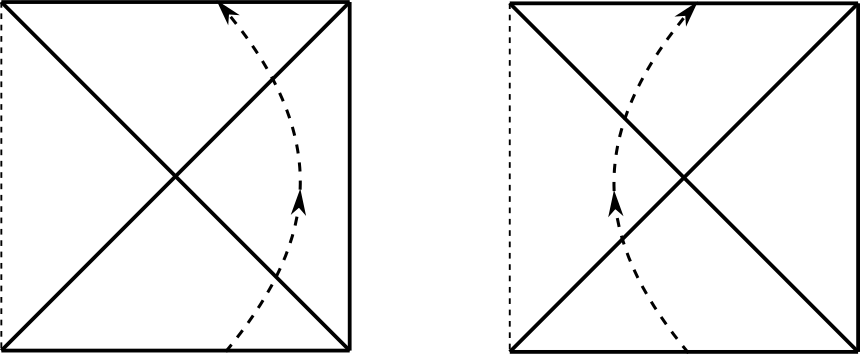}
	\caption{The bulk interpretation, the left diagram depicts a conventional boundary excitation (worldline completeted through time-reversal). The right diagram depicts the bulk worldline of the dressed unitary. Notice that is at all times undetectable by an observer living on the right boundary. Image recreated from a similar one in \cite{Papadodimas:2017qit}.}
	\label{behindthehorizon}
\end{figure}

Continuing in this line of development, in \cite{Papadodimas:2017qit} a particular out-of-equilibrium CFT state of the form 
\begin{equation}
|\Psi\rangle = e^{-\beta {\cal H}/2}U(\op) e^{\beta {\cal H}/2} |H\rangle
\label{nonequilibrium}
\end{equation} was considered, where $|H\rangle$ is a typical pure state defined in a small band of energy and $\op$ represents simple local operators of the CFT, centered at a particular time $t = t_0$. $U$ is a unitary operator constructed out of the simple operators. $\beta$ and $\cal H$ are the inverse temperature and the system Hamiltonian respectively. The unique feature of this state is that its non-equilibrium nature is not revealed when probed by a finite number of light operators, namely, $\frac{d}{dt}\langle \op_1 ...\op_n\rangle_{\Psi} = 0$. Here the subscript $\Psi$ denotes the fact that the expectation value is computed over the state $|\Psi\rangle$. However, if probed by the system Hamiltonian, $\frac{d}{dt}\langle {\cal H}\op_1 ...\op_n\rangle_{\Psi} \ne 0$, thereby demonstrating that this state is genuinely a non-equilibrium state. 
These states are therefore indistinguishable from the state $|H\rangle$ from the perspective of light operators. This feature is very different from a typical non-equilibrium state like $|\Psi\rangle_{\rm typical} = U(\op) |H\rangle$ for which $\frac{d}{dt}\langle \op_1 ...\op_n\rangle_{\Psi_{\rm typical}} = \frac{d}{dt}\langle {\cal H}\op_1 ...\op_n\rangle_{\Psi_{\rm typical}} = 0$. In holographic CFTs, these atypical non-equilibrium states correspond to black hole states with excitations localized behind the horizon \cite{Papadodimas:2017qit}. In other words, these excitations manifest themselves as particles with worldlines which, throughout their lifetime, are causally disconnected from the asymptotic boundary and hence cannot be detected by low-energy observers sitting on the boundary. A holographic visualization of such a state is provided in the right diagram of figure \ref{behindthehorizon}, while on the left we show the gravity dual of a typical non-equilibrium state of the form $|\Psi\rangle_{\rm typical}$.

The non-equilibrium state of the form \eqref{nonequilibrium} will constitute the basic set-up of the present work.
In \cite{Papadodimas:2017qit}, this was argued from the general expectation from statistical mechanics that a generic pure state like $|H\rangle$, in a large system, resembles a thermal state at the level of the resolution power of the light operators. The error term scales with $\frac{1}{S}$ or $e^{-S}$, depending on the ensemble, $S$ being the entropy of the system, which is parametrically large compared to the dimension of the small algebra mentioned above.   In this work we will forego the assumption, rather, we will demonstrate this very special indetectability property from first-principle computations, at the level of conformal blocks in the large-$c$ limit for ultra-high energy states. Conformal dimension $H$ of the CFT operator corresponding to such ultra-high energy states scales with the central charge, namely, $\sqrt{H/c}\gg 1$.

We will proceed in two different pathways, both eventually leading to the same conclusion. The first of the approaches exploits the uniformization map which we discuss through sections \ref{sec:KMS-HHKL}, \ref{sec:transformation-unitary} and \ref{sec:single-valued-geometry}. The uniformization map is a conformal transformation, which turns out to be extremely useful in expressing a heavy-light correlator of the form $\langle O_H O_H O_L \cdots O_L\rangle$ in terms of correlators of light operators up to some multiplicative factors coming from the decoupled heavy sector and conformal transformation rules \cite{Fitzpatrick:2015zha,fitzpatrick2016conformal, Banerjee:2018tut,Anous:2019yku}. In the literature, correlators of the above form are often schematically termed ``HHL...L correlators''. The uniformization method utilizes the heavy-light decoupling at the level of conformal blocks. In this work we will focus on the identity block contributions, which is a reasonable assumption for holographic CFTs. We will start section \ref{sec:KMS-HHKL} with a brief review of the uniformization method for computing the identity block of the HHLL correlator in the large central charge limit. We will demonstrate that in the regime when $\sqrt{H/c} \gg h_i$, $h_i$ being the scaling dimensions of the light operators, an effective notion of thermality emerges automatically from the identity block. 
Accordingly, from the emergent periodicity of the Euclidean time coordinate, one can precisely identify the temperature in the aforementioned regime. This has a simple CFT interpretation as this regime sets an effective characteristic length scale for the distance of the light operators in the correlation function, the limit $\sqrt{H/c} \gg h_i$ results in an OPE limit where the identity block dominates the conformal block decomposition. Therefore, in this limit the thermal properties of the full correlator are passed down to the identity block.

In section \ref{sec:transformation-unitary}, we show that in the same high-temperature regime, by the exact same mechanism, the unitary operators appearing in the state \eqref{nonequilibrium} simply drop out of the identity blocks which assume the general form
\begin{equation}
\langle H|e^{\beta {\cal H}/2}U^{\dagger} e^{-\beta {\cal H}/2}\, \op_1...\op_n\, e^{-\beta {\cal H}/2}U e^{\beta {\cal H}/2} |H\rangle_{I},
\end{equation} 
where $O_i$ denote the light operators.

In section \ref{sec:single-valued-geometry} we will provide a geometric picture of this method, which will in turn justify the applicability of the uniformization method in the two apparently uncorrelated applications discussed in section \ref{sec:KMS-HHKL} and section \ref{sec:transformation-unitary} respectively. Interestingly, we note that in the high-temperature regime mentioned above
the uniformization method results in a mapping to a general emergent thermal cylinder geometry. As a consequence, we find that for any finite number of light operators, the resulting Lorentzian correlator has to satisfy an emergent  Kubo-Martin-Schwinger (KMS) condition. 

In section \ref{sec:GNS}, we will follow the second pathway to reveal the defining feature of the non-equilibrium state \eqref{nonequilibrium}. This path is purely algebraic and exploits the vital features of the von Neumann algebra in axiomatic quantum field theory \cite{Haag, Witten:2018zxz}. In the case of holographic CFTs, these are the algebras of local boundary observers.
In the context of quantum gravity and its large $N$ approximation, this algebra of local observers, and in particular the classification thereof have experienced a resurgence of interest \cite{Leutheusser:2021frk, Leutheusser:2021qhd, Witten:2021jzq, Witten:2021unn, Sorce:2023fdx}. By identifying the algebra of observables at large $N$ as a type III algebra it is possible to utilize the machinery of Tomita-Takesaki theory to construct a set of half-sided modular inclusions that evolve a local operator beyond the event horizon.
We will exploit the properties of the emergent KMS-state in this section and will apply the GNS-construction for the following general linear form
\begin{equation}
\omega(\op) = \langle H|\op|H\rangle.
\end{equation} 
This will demonstrate the power of this construction in realizing the Lorentzian correlators in the CFT as time-ordered correlators in an auxiliary quantum system, namely the GNS-representatives of the light operators $\pi(\op_i)$. 
Furthermore, since $|H\rangle$ is time-translation invariant, we shall construct a formal Hermitian operator $\mathcal{K}$ (the modular Hamiltonian in the sense of \cite{Haag}) that annihilates the GNS-vacuum. 
We will show that the emergent KMS-condition automatically implies an order-reversing property from the modular Hamiltonian 
\begin{equation}
\exp(\beta \mathcal{K}/2) \pi(A) |K_0\rangle\rangle = |K_0 A\rangle\rangle = \pi_r(A^{\dagger})|K_0\rangle\rangle.
\end{equation}
Through this, using the expression of the modular Hamiltonian in terms of left- and right-representation of the Hamiltonian
\begin{equation}
\mathcal{K} = \pi(H) - \pi_r(H).
\end{equation}
we obtain the following identity
\begin{equation}
e^{-\beta \pi(H)/2} \pi(U) e^{\beta \pi(H)/2}|K_0\rangle\rangle = \pi_r(U)^{\dagger}|K_0\rangle\rangle,
\end{equation}
where $|K_0\rangle\rangle$ is the GNS-vacuum. The important realization of this identity is that the exponential factors map the representative of the unitary operator to the commutant algebra. Not only does the annihilation property simply follow as a consequence, but, it also clarifies the role of the latter in creating a bulk behind-the-horizon excitation. This will lead us to the same conclusion as obtained through the uniformization method, namely, the atypical non-equilibrium state \eqref{nonequilibrium} is actually indistinguishable from a typical equilibrium state when probed by a bunch of light operators. 
Finally we will conclude our paper in section \ref{sec:conclude} with some interesting discussions and future outlooks.

\section{The emergent thermality and the HHLL correlator}
\label{sec:KMS-HHKL}
As a warm-up for our original problem, in the present section, we will review the uniformization mechanism as applied to the HHLL out-of-time-ordered correlator in semi-classical CFT \cite{Fitzpatrick:2015zha}. Endowed with the machinery developed in this simple, yet non-trivial, setting we will apply the same algorithm in the following section to the non-equilibrium state \eqref{nonequilibrium}.

With this aim in view, to set up the stage, let us consider two light operators that are located at the same point in space but separated in time. In Lorentzian time, we consider expectation values on an energy eigenstate $|H\rangle$ of the form
\begin{equation}
\label{eqn:corr-H-Lorentzian}
A_L(t_1,t_2) = \langle H| {\cal O}(t_2){\cal O}(t_1)|H\rangle,
\end{equation}
under the usual semiclassical assumptions, i.e. $c\gg 1$ with $H > c/24$ and where the scaling weight $h$ of the operator $\cal O$ is ${O}(c^{0})$. For simplicity, we have set the angular position of the operator insertions as $\phi = 0$. Unless otherwise stated, we will use this configuration throughout the paper.

Our strategy will be to consider the problem on the Euclidean plane from which, we will be able to obtain the Lorentzian correlator \eqref{eqn:corr-H-Lorentzian} through an appropriate analytic continuation. For this purpose, we first write \eqref{eqn:corr-H-Lorentzian} in the Heisenberg form
\begin{equation}
A_L(t_1,t_2) = \langle H|e^{i{\cal H}t_2}{\cal O}e^{i{\cal H}(t_1-t_2)}{\cal O}e^{-i{\cal H}t_1}|H\rangle.
\end{equation}
The edge-of-the-wedge theorem allows us to analytically continue $t=(t_1-t_2)$ to the upper half-plane (see section \ref{sec:single-valued-geometry} for details on time-ordering and analytic continuation). After Euclideanizing, the state $|H\rangle$ is taken to be a primary state of the CFT. Separating the light operators by appropriate imaginary parts, results in the following Euclidean correlator on the cylinder
\begin{equation}
A(\tau) = \langle {\cal O}_H(\infty) {\cal O}(0) {\cal O}(\tau) {\cal O}_H(-\infty)\rangle,
\end{equation}
Furthermore, with the exponential map $z=\exp(\tau +i\phi)$, the correlator takes the form 
\begin{equation}
\label{eqn:Atau}
A(\tau) = \langle {\cal O}_H(\infty) {\cal O}(1){\cal O}(e^{\tau}){\cal O}_H(0)\rangle.
\end{equation}
At this point, we apply the uniformization method to extract the leading-order contribution in the large-$c$ regime from this Euclidean correlator. 
This uniformization method exploits the heavy-light decoupling in the limit of a large central charge, which simplifies the evaluation of a mixed heavy-light correlator to that of a light correlator on a background geometry determined by the heavy sector. This method therefore turns out to be an extremely important tool for computing the expectation values of light operators on heavy states.
Let us now quickly review the basic algorithm for readers unfamiliar with this method. For further details, interested readers are referred to \cite{Fitzpatrick:2015zha,fitzpatrick2016conformal, Banerjee:2018tut} and in particular \cite{Anous:2019yku} which specifically deals with heavy-light correlators of the HHL...L form. 

As mentioned above, the uniformization method exploits the factorization of the heavy and light sectors of 2d CFT conformal blocks in the large central charge regime \cite{Zamolodchikov:1987avt} expressed in the form
\begin{equation}
\langle {\cal O}_L(z_1)...{\cal O}_L(z_n) {\cal O}_H(x_1)...{\cal O}_H(x_m)\rangle_{V} = Q(z_i; x_i) e^{-\frac{c}{6} f(x_i)},
\label{HeavyLightDecoupling}
\end{equation}
where the subscript $V$ indicates a set of exchanged conformal representations and 
\begin{equation}
\label{eqn:decoup-exp}
\langle {\cal O}_H(x_1)...{\cal O}_H(x_m)\rangle_{V} = e^{-\frac{c}{6} f(x_i)}.
\end{equation}
Both the functions $Q(z_i; x_i)$ and $f(x_i)$ are assumed to scale as $c^{0}$ in the large central charge regime. The essential statement is that the light operators are sensitive to the presence of the heavy operators but the dynamics of the heavy sector is unaffected by the light operators. In other words, the light operators can be thought of as a probe in the background created by the heavy operators.

Let us write down the conformal block explicitly by inserting a projector that sums up the contributions of intermediate states given by a primary state and its descendants. 
Furthermore, let us assume that the dominant contribution to the conformal block comes from the identity Virasoro block. Although this can be rigorously proved only for a few very special cases \cite{Hartman:2013mia, Hartman:2014oaa}, one would expect this to hold for holographic CFTs \cite{El-Showk:2011yvt} having a well-defined large-c limit and sparse spectrum for low-lying operators. This expectation follows from the decoupling limit of the heavy contribution in the form of an exponential as in \eqref{eqn:decoup-exp}. This form motivates to run a saddle-point analysis that in turn justifies this expectation for holographic CFTs \cite{Anous:2019yku}. we can write
\begin{align}
\label{eqn:projector}
& \langle {\cal O}_L(z_1)...{\cal O}_L(z_n) {\cal O}_H(x_1)...{\cal O}_H(x_m)\rangle_{0} \nonumber \\
& = \sum_{\{n\}}\frac{\langle {\cal O}_H {\cal O}_H L_{-n_1}...L_{-n_m}|0\rangle \langle 0|L_{n_m}...L_{n_1}{\cal O}_L...{\cal O}_L\rangle}{\langle 0|L_{n_m}...L_{n_1}L_{-n_1}...L{-n_m}|0\rangle},
\end{align}
where the sum over $\{n\}$ crudely indicates the sum over all ordered lists of integers of indefinite length $(n_1, n_2,...)$ with their values starting from $n_i = 2$ for the identity block, this by virtue of the fact that the vacuum state is annihilated, $L_n|0\rangle =0 $, for all $n> -2$. 

%The last fact follows from the definition of identity operator for which $h = {\bar h} = 0$ with its associated state identified as the vacuum $|0\rangle$. This state, by definition, transforms trivially under the global conformal group. Furthermore, by virtue of it being a primary state, the vacuum gets annihilated by all the generators $L_p$ for $p>0$. Therefore, the projector can only contain generators $L_m$ with $m \le -2$ for the identity block.

From \eqref{eqn:projector} it is not evident whether the large-$c$ regime  provides any obvious simplification. This is because although the denominator scales with positive powers of $c$ \cite{Fitzpatrick:2014vua}, the heavy sector in the numerator can potentially produce the same power of $c$ following the relation
\begin{equation}
\left[L_n, {\cal O}_H(z)\right] = H(1+n) z^n {\cal O}_H(z) + z^{n+1} \partial_z {\cal O}_H(z), 
\end{equation}
where, as mentioned earlier, the scaling dimension of the heavy operator, $H$ scales with $c$.

In order to resolve this problem, following the proposal advocated in \cite{Fitzpatrick:2015zha}, we consider a transformation to a different conformal frame $z\rightarrow z(w)$ and subsequently insert the projector from before
\begin{align}
& \langle {\cal O}_L'(w(z_1))...{\cal O}_L'(w(z_n)) {\cal O}_H'(w(x_1))...{\cal O}_H'(w(x_m))\rangle_{0} \nonumber \\
& = \sum_{\{n\}}\frac{\langle {\cal O}_H' {\cal O}_H' L_{-n_1}...L_{-n_m}|0\rangle \langle 0|L_{n_m}...L_{n_1}{\cal O}_L'...{\cal O}_L'\rangle}{\langle 0|L_{n_m}...L_{n_1}L_{-n_1}...L{-n_m}|0\rangle}.
\end{align}
The Virasoro generators $L_n$ are defined as the coefficients of the Laurent series expansion of the stress-energy tensor which yields
\begin{equation}
L_n = \frac{1}{2\pi i} \oint dz\, z^{n+1}T(z),
\end{equation}
 With this, the domination of the identity block can be restated as the domination of the OPE channel in the correlator, where the unit operator and all its descendants, namely, $T, T^n, \partial T, T^m \partial T$ etc. run along the internal lines.
From the perspective of dual gravity theory in AdS$_3$, the basis state appearing in the projector \eqref{eqn:projector}, namely a state of the form $L^{k_1}_{-n_1}...L^{k_m}_{-n_m}|0\rangle$ can accordingly be interpreted as a $k$-graviton state with $\sum_\alpha k_\alpha = k$.
 Focusing our attention on the heavy factor in the numerator for a moment, we can construct a set of generating functions $F^{(n)}(w_1,...,w_n; x_i, H)$ labeled by an integer $n$ for every heavy factor that contains $n$ Virasoro generators
\begin{equation}
F^{(n)}(w_1,...,w_n; x_i, H_i) = \langle T(w_1)...T(w_n){\cal O}_H(x_1)...{\cal O}_H(x_m)\rangle.
\end{equation}
Let us consider the generating function $F^{(1)}(w; x_i, H_i)$ which in the $z$-conformal frame reads
\begin{align}
& F^{(1)}(z; x_i, H) = \left(\frac{\partial w}{\partial z}\Big\vert_{x_1}\right)^{-H_1}... \left(\frac{\partial w}{\partial z}\Big\vert_{x_m}\right)^{-H_m} \left(\frac{\partial w}{\partial z}\right)^{-2} \nonumber \\
& \times \langle \left(T(z) - \frac{c}{12}S[w,z]\right) {\cal O}_H(x_1)...{\cal O}_H(x_m)\rangle,
\end{align}
where $S[w,z]$ is the Schwarzian derivative
\begin{equation}
\label{eqn:Schwarzian}
S[w,z] = \frac{w'''(z)}{w'(z)} - \frac{3}{2}\left(\frac{w''(z)}{w'(z)}\right)^2,
\end{equation} 
which arises due to the transformation of the stress-energy tensor $T(w)$ under the conformal transformation $z\rightarrow w$.
Generically, this factor will be proportional to $c$, but if we choose a fine-tuned conformal frame that solves the ODE
\begin{equation}
S[w,z] = \frac{12}{c}T_H(z), \;\;\;\;\; T_H(z) = \langle T(z) {\cal O}_H(x_1)...{\cal O}_H(x_m)\rangle,
\end{equation}
then the generating function, and hence all terms involving a single Virasoro generator vanish. In addition, one can show that all other terms that involve greater number of Virasoro generators have their growth in the large-$c$ limit retarded by a single power of $c$. As a consequence, in this conformal frame, the vast majority of terms are suppressed by inverse of powers of $c$. The only term that remains is the projection onto the vacuum state \cite{Fitzpatrick:2015zha}. Hence, in this frame, only the vacuum state survives the large-$c$ limit as an intermediate state. This yields the factorization
\begin{align}
& \langle {\cal O}_L'(w(z_1))...{\cal O}_L'(w(z_n)) {\cal O}_H'(w(x_1))...{\cal O}_H'(w(x_m))\rangle_{0} \nonumber \\
& = \langle {\cal O}_L'(w(z_1))...{\cal O}_L'(w(z_n))\rangle \langle {\cal O}_H'(w(x_1))...{\cal O}_H'(w(x_m))\rangle.
\end{align} 
Transforming the full expression back to the original $z$-frame and applying \eqref{HeavyLightDecoupling} results in 
\begin{align}
& \langle {\cal O}_L'(w(z_1))...{\cal O}_L'(w(z_n))\rangle\, \langle {\cal O}_H'(w(x_1))...{\cal O}_H'(w(x_m))\rangle. \nonumber\\
& = \left(\frac{\partial w}{\partial z}\Big|_{x_1}\right)^{-H_1}...\left(\frac{\partial w}{\partial z}\Big|_{x_m}\right)^{-H_m}\left(\frac{\partial w}{\partial z}\Big|_{z_1}\right)^{-h_1}...\left(\frac{\partial w}{\partial z}\Big|_{z_m}\right)^{-h_m} Q(z_i; x_i, H_i)e^{-\frac{c}{6}f(x_i)}
\end{align}
Stripping of all factors independent of the light sector we can solve for $Q(z_i; x_i, H_i)$
\begin{equation}
Q(z_i; x_i, H_i) = \left(\frac{\partial w}{\partial z}\Big|_{z_1}\right)^{h_1}...\left(\frac{\partial w}{\partial z}\Big|_{z_m}\right)^{h_m} \langle {\cal O}_L'(w(z_1))...{\cal O}_L'(w(z_n))\rangle.
\end{equation}
Hence we conclude that the computation of the identity block contribution of the expectation value of a set of light operators on a heavy state is given by a vacuum expectation value of the light sector dressed with some derivative factors in a very specific conformal frame that depends on the heavy sector.

%%%%%%%%%%%%%%%%%%%%%%%%%%%%%%%%%%%%%%%%%%%%%%%%%%%%%

Applying this method in our setup \eqref{eqn:Atau}, yields the identity block contribution to the correlator\footnote{We only partially fix the conformal channel by restricting to exchange of the vacuum representation between the collective heavy sector and the light sector, in the preceding section we left the remainder of the channel ambiguous as the choice does not affect the discussed method. In the actual implementation in section 3 we will argue that we are implicitly projecting out the lowest-lying conformal block of the light sector.}, to leading order in $1/c$ as
\begin{equation}
A_{I}(\tau) = \left(\frac{\partial w}{\partial z}\Big|_{1}\right)^h \left(\frac{\partial w}{\partial z}\Big|_{e^{\tau}}\right)^h \left( w(1)-w(e^{\tau})\right)^{-2h},
\end{equation}
where
\begin{equation}
S[w,z] = \frac{12}{c} T_H(z), \;\;\; {\rm with} \;\;\; T_H(z) = \langle T(z) {\cal O}_H(\infty) {\cal O}_H(0) \rangle.
\label{SchwarzianODE}
\end{equation}
Here $S[w,z]$ is the Schwarzian derivative given in \eqref{eqn:Schwarzian}.
In the particular case of just two heavy operators, the heavy stress tensor expectation value is fixed by conformal symmetry and is given by
\begin{equation}
T_H(z) = \langle T(z) {\cal O}_H(\infty) {\cal O}_H(0) \rangle = \frac{H}{z^2}.
\end{equation}
As a result, the Schwarzian equation \eqref{SchwarzianODE} has a simple solution 
\begin{equation}
\label{eqn:coord-change}
w(z) = z^{i\alpha}, \;\;\;\;\; \alpha = \sqrt{24 H/c -1}.
\end{equation}
Note that we have $H>c/24$. This results in 
\begin{align}
A_{I}(\tau) = (i\alpha)^{2h} \frac{e^{\tau h(i\alpha-1)}}{(1-e^{i\tau\alpha})^{2h}} 
= \left(2\alpha \right)^{-2h} e^{-\tau h} \sin\left(\frac{1}{2}\tau \alpha\right)^{-2h}.
\end{align}
After stripping off the irrelevant factors this results in
\begin{equation}
A_I(\tau) \propto e^{-\tau h} \sin\left(\frac{1}{2}\tau \alpha\right)^{-2h}.
\label{2ptblock}
\end{equation}
This is almost periodic in Euclidean time with period $\tau\sim \tau + \frac{4\pi}{\alpha}$, i.e. thermal. The absolute thermality is of course spoiled by the overall exponential prefactor. This is, however, not a problem, since this can simply be interpreted as a consequence of the fact that $A_I(\tau)$ is only the identity block contribution to the full correlator \cite{Fitzpatrick:2015zha}. In the case where the full correlator would be truly thermal\footnote{As could be speculated from AdS$_3$/CFT$_2$, given a stress tensor of the form $T(z)=H/z^2$, mapping back to the Lorentzian cylinder and taking the resulting periodic functions (both the holomorphic and anti-holomorphic sector) as input for the Ba\~nados geometry results in a black hole with inverse temperature \eqref{eqn:temperature} if $\bar{H}=H>c/24$. \cite{Banados:1998gg}}, including the full conformal block expansion should restore the periodicity $\tau\sim \tau + \frac{4\pi}{\alpha}$ and we obtain a thermal expectation value with temperature\footnote{One might wonder why the inverse temperature is only half of the periodicity in $\tau$ in \eqref{2ptblock}. This is essentially because of the holomorphic/anti-holomorphic decoupling. Through the inverse exponential map the Euclidean time is related to the radial plane coordinates through $\tau_E=\frac{1}{2}\log(z\bar{z})$ a shift $z\rightarrow e^{\eta} z$ therefore results in only half the shift in the physical Euclidean time $\tau_E \rightarrow \tau_E + \frac{1}{2}\eta$. In section 4.1 we avoid this issue by introducing $\tilde{\tau}, \tilde{\phi}$-coordinates for the uniformized geometry.}
\begin{equation}
\label{eqn:temperature}
T= \frac{\alpha}{2\pi}.
\end{equation}
In the regime where the temperature $T$ is very large compared to the scaling weight $h$, there is an OPE limit for small values of $\tau$ where the contribution of the sine factor dominates over the exponential factor and the approximate correlator is effectively periodic in $\tau$. This expresses the fact that in this regime the full correlator is increasingly accurately approximated by the identity block\footnote{In addition to being periodic in $\tau$, in this limit the conformal block in fact satisfies the KMS-condition, see section 4.3. Due to the identical nature of the two light-operators the operator ordering swtich is not obvious, but when $\tau$ is analytically continued it becomes clear that there is 
a branch cut in the Lorentzian regime that signals a change of ordering.}.

For the sake of completeness, one can perform the exact same computation where the operators are not presumed to be located at $\phi_1=\phi_2 =0$, keeping $\phi_1$ and $\phi_2$ general results in
\begin{equation}
A_{I}(\tau,\phi_1,\phi_2) \propto e^{-ih(\phi_1+\phi_2)}e^{-h\tau}\sin\left(\frac{1}{2}\alpha \tau -i\alpha(\phi_1-\phi_2)\right)^{-2h}.
\end{equation}
From this, we can conclude that all previously derived conclusions do not depend on the light operators being located at the same point in space.

\section{A transformation rule for the unitary operators}
\label{sec:transformation-unitary}
This brings us towards the main claim of this paper. {\it We will now show that the same uniformization mechanism that produced the thermal identity block in the high-temperature regime is also responsible for hiding a particular class of dressed unitary operators. In the $\sqrt{H/c}\gg h$ regime this class of operators coincides with dressed the unitary operators in the non-equilibrium state \eqref{nonequilibrium}.} In order to demonstrate this, let us consider an arbitrary expectation value of light operators on the non-equilibrium state \eqref{nonequilibrium} on the Lorentzian cylinder, i.e.
\begin{equation}
\langle H| e^{\beta {\cal H}/2} U^{\dagger}(0)e^{-\beta H/2} {\cal O}(z_1)...{\cal O}(z_n)e^{-\beta H/2}U(0)e^{\beta H/2}|H\rangle.
\end{equation}
Following the discussion of the previous section, we will consider an ``equivalent'' problem on the Euclidean radial plane instead of the Lorentzian cylinder\footnote{see section \ref{sec:single-valued-geometry} for a more detailed discussion regarding the time-ordering and analytic continuation from Euclidean to Lorentzian correlators}. 
\begin{equation}
\langle {\cal O}_{H}(\infty)e^{\beta {\cal H} /2} U^{\dagger}(1) e^{-\beta {\cal H}/2} {\cal O}...{\cal O} e^{-\beta {\cal H}/2}U(1)e^{\beta {\cal H}/2}{\cal O}_{H}(0)\rangle.
\end{equation}
In this case, by ``equivalent'' we mean that one can retrieve the correctly ordered Lorentzian correlator through appropriate analytic continuation of the Euclidean times to imaginary values.
It is worth noting that in the original formulation the transformation $e^{-\beta H/2}$ acts as a shift by $\beta /2$ in Euclidean time. On the radial plane $z=e^{\tau+i\phi}$, hence $\tau \rightarrow \tau + \beta /2$ leads to $z \rightarrow e^{\beta/2}z$. Our correlator can therefore equivalently be written as
\begin{equation}
\langle {\cal O}_{H}(\infty)U^{\dagger}(e^{-\beta /2}) {\cal O}...{\cal O} U(e^{\beta /2}){\cal O}_{H}(0)\rangle.
\end{equation}
This correlator is Euclidean and none of the operators are (assumed to be) located at the same point. Hence, all operators commute with each other and we can rewrite it into a more suggestive form
\begin{equation}
\langle {\cal O}_H(\infty){\cal O}_{H}(0)U^{\dagger}(e^{-\beta /2})U(e^{\beta /2}){\cal O}...{\cal O}\rangle.
\end{equation}

To apply the uniformization map, we must first establish a transformation rule for the unitary operators under conformal transformations. Let us assume for simplicity that our unitary is given by an exponential of a Hermitian scalar primary\footnote{A typical example of this is quenching the system with a source for a scalar primary.}.
\begin{equation}
U(z) = e^{i\op(z)}.
\end{equation}
In this case, constructing the the transformation rule of $U(z)$ from the transformation rule of $O(z)$ results in 
\begin{equation}
U'(w) = e^{i\op(z)\left(\frac{\partial z}{\partial w}\right)^h} \equiv e^{i \op'(w)}.
\end{equation}
Going to the uniformization coordinates \eqref{eqn:coord-change} $w=z^{i\alpha}$, this yields 
\begin{equation}
U'\left(e^{i\alpha \beta/2}\right) = e^{ i \op\left(e^{\frac{\beta}{2}}\right)(i\alpha)^{h}\left(e^{i\alpha \beta /2}\right)^{h}e^{-\beta h/2}}.
\end{equation}
The goal of this section will be to demonstrate that this operator has a special property when the real parameter $\beta$ equals the inverse of the temperature given in \eqref{eqn:temperature}. Hence we will Eliminate $\beta$ using the relation \eqref{eqn:temperature}, in which case we obtain
\begin{equation}
U'\left(e^{i\pi}\right) = \exp\left(i\alpha^{-h}e^{-i\frac{3}{2}\pi h}e^{\pi h/\alpha}\op\left(e^{\beta/2}\right)\right)
\label{unitary1}
\end{equation}
Similarly for the adjoint unitary operator we find
\begin{equation}
U'^{\dagger}\left(e^{-i\pi}\right) = \exp\left(-i\alpha^{-h}e^{i\pi h/2}e^{-\pi h/\alpha}\op\left(e^{-\beta/2}\right)\right)
\label{unitary2}
\end{equation}
Rearranging the expressions \eqref{unitary1} and \eqref{unitary2} and rotating the complex plane by a full $2\pi$ yields\footnote{Note that this implies we are also rotating all the light operators, this will result in a set of multiplicative factors. Since in the application we have in mind the multiplicative factors appear on both sides of the equation and hence divide out we will not bother to write these factors down explicitly.}
\begin{align}
&\exp\left(i e^{i\pi h/2} \op'\left(e^{i\pi}\right)\right) = \exp\left(i\alpha^{-h}e^{\pi h/\alpha}\op\left(e^{\beta/2}\right)\right)\\
&\exp\left(-i e^{i\pi h/2} \op'\left(e^{i\pi}\right)\right) = \exp\left(-i\alpha^{-h}e^{-\pi h/\alpha}\op\left(e^{-\beta/2}\right)\right)
\end{align}
As a final step, for reasons that will become clear very soon, we rescale the operators $\mathcal{O}$ by an overall factor
\begin{equation}
\tilde{\op} \equiv \alpha^{-h}\op\,,
\end{equation}
which results in the expressions
\begin{align}
&\exp\left(i e^{i\pi h/2} \alpha^h \tilde{\op}'\left(e^{i\pi}\right)\right) = \exp\left(ie^{\pi h/\alpha}\tilde{\op}\left(e^{\beta/2}\right)\right)\\
&\exp\left(-i e^{i\pi h/2} \alpha^h \tilde{\op}'
\left(e^{i\pi}\right)\right) = \exp\left(-ie^{-\pi h/\alpha}\tilde{\op}\left(e^{-\beta/2}\right)\right)
\end{align}

We are now in a position to run the uniformization algorithm sketched in the previous section. Going to uniformizing $w$-coordinates \eqref{eqn:coord-change},
\begin{align}
&\langle \exp\left(-i e^{i\pi h/2} \alpha^h \tilde{\op}'\left(e^{i\pi}\right)\right) \op_1'(w_1)...\op_n'(w_n)\exp\left(i e^{i\pi h/2} \alpha^h \tilde{\op}'
\left(e^{i\pi}\right)\right)\rangle \nonumber\\
&=\left(\frac{\partial z}{\partial w_1}\right)^{h_1}...\left(\frac{\partial z}{\partial w_n}\right)^{h_n} \times \nonumber\\ 
&\langle H|\exp\left(-ie^{-\pi h/\alpha}\tilde{\op}\left(e^{-\beta/2}\right)\right)\, \op(z_1)...\op(z_n)\exp\left(ie^{\pi h/\alpha}\tilde{\op}\left(e^{\beta/2}\right)\right)|H\rangle_I,
\end{align}
this procedure exclusively projects out the identity block\footnote{Note on terminology, we have been referring to this block as the identity block, but the only part of the OPE channel that is fixed is the contraction between the two heavy operators where we restricted to the exchange of the vacuum representation. Due to the implicit kinematics of our light operators, we project out the representation with the lowest conformal weight from the other OPEs. This will only be the identity block if we arbitrarily demand that the light operators are pairwise identical. Since there is no need for such a restriction perhaps the `lowest-lying block' would be a more appropriate term.}. In order to keep this fact in mind we append the subscript $I$. Two things are worth noting at this point. First, the heavy sector has been factored out of the left-hand side and second, the unitary operators cancel each other, also on the left-hand side.
This results in the following expression
\begin{align}
\label{result-1}
&\langle H|\exp\left(-ie^{-\pi h/\alpha}\tilde{\op}\left(e^{-\beta/2}\right)\right)\, \op(z_1)...\op(z_n)\exp\left(ie^{\pi h/\alpha}\tilde{\op}\left(e^{\beta/2}\right)\right)|H\rangle_I  \nonumber \\
&=\left(\frac{\partial z}{\partial w_1}\right)^{-h_1}...\left(\frac{\partial z}{\partial w_n}\right)^{-h_n}  \langle  \op_1'(w_1)...\op_n'(w_n)\rangle,
\end{align}
where the prefactors are evaluated from \eqref{eqn:coord-change}. To reflect on what we found, it appears that a very specific combination of a dressed unitary operator multiplied with an additional exponential prefactor disappears entirely from the identity block in the large-$c$ regime, leaving the identity block in the exact same form it would have been if it had not been there from the start.

Equation \eqref{result-1} possesses an interesting interpretation as follows. In \cite{Papadodimas:2017qit}, the state \eqref{nonequilibrium} was interpreted as an atypical bulk black hole state with a non-equilibrium excitation localized behind the horizon. Note that in the same high-temperature limit as in the HHLL computation above, i.e. $\alpha \gg h$, the additional multiplicative exponential prefactors reduce exactly to unity. As a consequence, in this limit, where the identity block converges towards the full correlation function, the behind-the-horizon non-equilibrium operators become invisible to the light operator algebra! The take-away message is {\it through the machinery of uniformization, we show that we can corroborate this interpretation in the large-$c$ regime, but most importantly, without having to approximate the heavy eigenstate as an effective thermal ensemble.}

\section{Emergent single-valuedness in the high temperature regime}
\label{sec:single-valued-geometry}
Computations in the previous two sections revealed the power of the uniformization map in  reducing the computation of the identity block of a heavy-light correlator to a computation of the full
correlator over only the light sector on a particular conformal frame up to some
derivative factors.
In the present section, we will argue that the underlying mechanism towards obtaining the effective thermal physics in the two cases, discussed subsequently in the previous two sections, is actually the same and follows directly from an emergent single-valuedness in the $w$-coordinates around the periodic $\tau$-coordinate in the regime where $\alpha\gg h$. This realization in turn provides a geometric interpretation of the perturbative approach to behind-the-horizon physics.

In order to investigate the fate of periodicity through the uniformization map, let us first note that the coordinate transformation
\begin{equation}
w=z^{i\alpha},
\end{equation}
itself is manifestly periodic under $z\rightarrow e^{\frac{2\pi}{\alpha}}z$. The light correlator on the $w$-plane is single-valued under this periodicity. The derivative factors on the other hand are not manifestly single-valued. This can be explicitly checked by evaluating this factor under the afore-mentioned coordinate transformation as
\begin{equation}
\label{eqn:derivatives}
\left(\frac{\partial w}{\partial z}\right)^h = \left(\frac{\partial z}{\partial w}\right)^{-h}
=\left(i\alpha\right)^h w^h w^{\frac{ih}{\alpha}}.
\end{equation}
Clearly, the last factor $w^{\frac{ih}{\alpha}}$ is not single-valued under this cycle. Nevertheless, From \eqref{eqn:derivatives}, it is also obvious that the single-valuedness is nevertheless preserved in the regime $\alpha \gg h$. In what follows we provide a geometric interpretation of this reemergence.  

\subsection{Uniformizing geometry}
To gain a better intuition for the revival of periodicity in the $\alpha \gg h$ regime, let us resort to the geometric description of the uniformization map. Our initial domain is the radial plane with the line element
\begin{equation}
\label{eqn:z-coord}
ds^2 = \frac{1}{2} dz d\bar{z}, \;\;{\rm with}\;\; z = e^{\tau + i\phi}, \;\;\; \bar{z}= e^{\tau -i\phi},
\end{equation}
$\tau$ and $\phi$ being the usual cylinder coordinates. Let us now make the change of coordinates using the uniformization map\footnote{Note the relative complex conjugation of the exponents. Both the provided solution and the solution with the exponent complex conjugated solve the uniformization equation, but only for this combination of solutions is the resulting metric manifestly real-valued.}
\begin{equation}
w = z^{i\alpha}, \;\;\;\;\;\; \bar{w}=\bar{z}^{-i\alpha}.
\end{equation}
With this, the line element transforms to

\begin{align}
\label{eqn:line-element}
ds^2 = g_{\mu\nu}(w,\bar{w})\; dw^{\mu}\, dw^{\nu}
=\frac{1}{2 \alpha^2} w^{\frac{1}{i\alpha}-1}\bar{w}^{\frac{1}{-i\alpha}-1} \; dw \, d\bar{w}.
\end{align}
In order to visualize the resulting geometry, let us undo the exponential map from the $w$-coordinate chart by introducing the coordinates $\tilde{\tau}$, $\tilde{\phi}$ through
\begin{equation}
w=e^{\tilde{\tau}+i\tilde{\phi}}, \;\;\;\;\;\;\;\; \bar{w}=e^{\tilde{\tau}-i\tilde{\phi}},
\end{equation}
such that, in these new coordinates, the line element \eqref{eqn:line-element} takes the form
\begin{align}
\label{eqn:lineelement-tilde}
ds^2 =\frac{1}{\alpha^2}e^{\frac{2\tilde{\phi}}{\alpha}}\left(d\tilde{\tau}^2 + d\tilde{\phi}^2\right).
\end{align}

\begin{figure}
	\centering
		\includegraphics[scale=0.5]{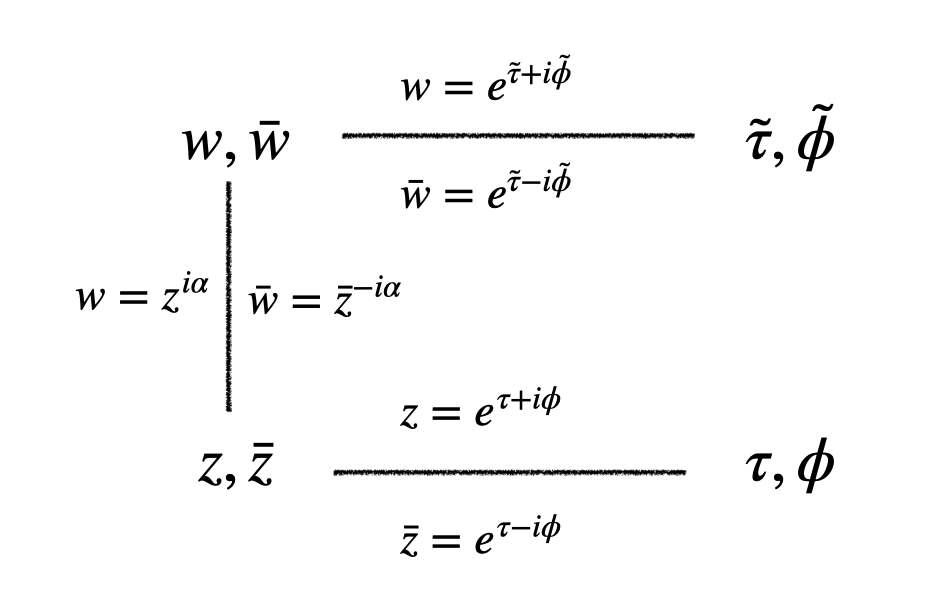}
	\caption{The internal relationships of the coordinate systems}
	\label{coordchange}
\end{figure}
The relation between all the coordinates discussed so far is summarized in figure \ref{coordchange}.
This transformation of coordinates makes it simple to recast the global periodicity condition in the original $\{\tau, \phi\}$ coordinates,
\begin{equation}
\phi \sim \phi + 2\pi.
\end{equation}
in terms of the new $\{\tilde\tau, \tilde\phi\}$ coordinates. By recalling that the $\{\tau, \phi\}$ coordinates are defined in terms of the $z$-coordinates through \eqref{eqn:z-coord}, 
the periodicity condition can be translated into
\begin{equation}
w=z^{i\alpha} = \left(e^{\tau + i\phi}\right)^{i\alpha} \sim \left(e^{\tau + i\phi +2\pi i}\right)^{i\alpha} = w e^{-2\pi \alpha}.
\end{equation} 
or in short $w\sim w e^{- 2\pi \alpha}$. In $\tilde{\tau},\tilde{\phi}$-coordinates this results in the periodicity condition
\begin{equation}
\tilde{\tau}\sim \tilde{\tau} + 2\pi \alpha.
\label{periodicity}
\end{equation}

\begin{figure}[h]
	\centering
 \includegraphics[scale=0.35]{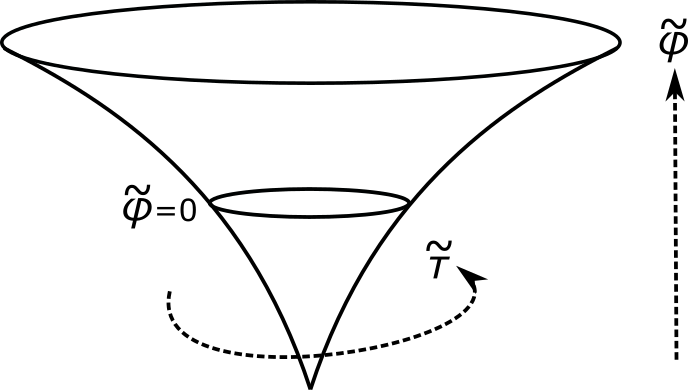}
	\caption{A embedding of the uniformizing geometry in three-dimensional space.}
	\label{blackhole}
\end{figure}

The resulting geometry is sketched in figure \ref{blackhole}.
%\footnote{A curious reader might wonder why this geometry is not identical to the usual `plumbing fixture' wormhole geometry. The reason is that the wormhole configuration is obtained when the coordinate change $w=z^{i\alpha}$ functions as a pull-back on the hyperbolic disk with constant negative curvature. On the contrary, we start with the flat cylinder, which through the same mapping yields figure \ref{blackhole}.}

The periodicity condition is useful, but it obscures some of the physics of the geometry due to the prefactor in the metric. At $\phi=0$ where the operators are inserted, the actual geodesic length of the cycle $x(s)$ for $0\leq s\leq 2\pi \alpha$ is given by
\begin{equation}
\label{eqn:length-scale}
L = \int_{0}^{2\pi \alpha} \frac{ds}{\alpha^2} = \frac{2\pi}{\alpha}.
\end{equation}
This will provide a very relevant scale in the high-temperature regime as we discuss next.

\subsection{The $\alpha \gg h$ regime}
We will now investigate the regime where $\alpha \gg h$ for the geometry \eqref{eqn:lineelement-tilde}. It is worth recalling that in this regime only, we found that the identity block of a general HHL...L correlator becomes single-valued under full rotations of the $\tilde{\tau}$-coordinate.
However, we should also keep in mind that in the same regime, a length scale emerges (relative to the circumference of the unit circle) which is given by \eqref{eqn:length-scale}. This immediately tells us that for the uniformization map to make sense in the examples elaborated in the previous two sections, one needs to implicitly assume that the light operators cannot be separated from each other by more than ${O}(1)$ multiples of $\frac{2\pi}{\alpha}$. Keeping the latter fact in mind, we 
restrict the coordinate range of $\tilde{\phi}$ through a perturbative expansion $\tilde{\phi} \rightarrow \tilde{\phi}_0 + \frac{\tilde{\phi}}{\alpha} + {O}((\tilde{\phi}/\alpha)^2)$. With this, the line element \eqref{eqn:lineelement-tilde} takes the form,
\begin{equation}
ds^2 = \frac{1}{\alpha^2} e^{\frac{2\tilde{\phi}_0}{\alpha}} (d\tilde{\tau}^2 + d\tilde{\phi}^2) + {O}\left(\left(\frac{\tilde{\phi}}{\alpha}\right)^2\right),
\end{equation}
which at the leading order can be identified as the metric of a dilated flat cylinder in 2D.

\begin{figure}[h]
	\centering
		\includegraphics[scale=0.35]{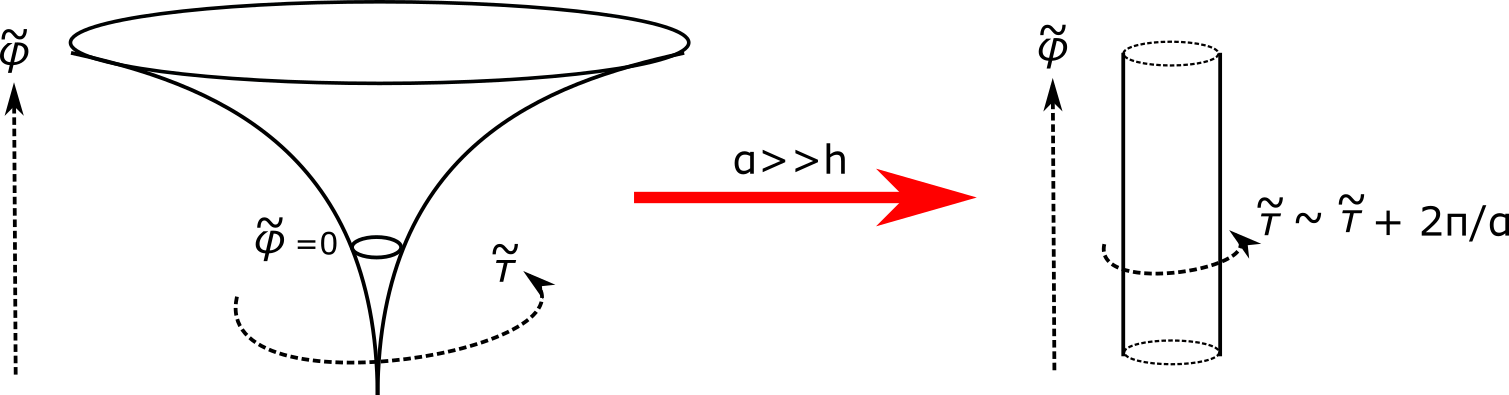}
	\caption{As a consequence of the regime where $\alpha$ is very large the variation of the width of the uniformizing geometry is small over variations of $\phi$ of the order of the periodicity in $\tau$. If we similarly restrict the insertions of the light operators to separations at the same scale then the light operators effectively live on an infinite thin cylinder.}
	\label{EffectiveCylinder}
\end{figure}

At this linearized order, if we introduce the unitary operators at $\tilde{\phi}_0=0$, the geodesic length of the compact dimension is $L=2\pi/\alpha$ which defines the actual $\tilde\tau$-periodicity for this configuration.

This brings us to the following conclusion. The regime where $\alpha$ becomes parametrically large compared to all the light operator scaling weights, the correlation function displays three separate but simultaneous effects:
\begin{enumerate}
\item OPE kinematics projects the lowest-lying conformal block out of the full HHL...L correlator.
\item This conformal block becomes a single-valued function of the $w_i$ coordinates under a full rotation around the symmetry axis.
\item The geometry described by the $w$-coordinates becomes that of an extremely thin flat cylinder as demonstrated, pictorially, through figure \ref{EffectiveCylinder}.
\end{enumerate}
In the next part of our story, we will show that the aforementioned three facts combine to form the remarkable conclusion in the Lorentzian regime, namely, every time-ordered correlator of the general form
\begin{equation}
\langle H|O_n(t_n)...O_1(t_1)|H\rangle
\end{equation}
satisfies the KMS-condition in the regime $\alpha\gg h$.

\subsection{The emergent KMS-condition}
\label{sec:emergentKMS}
The KMS condition is a consequence of thermality that functions as a periodicity condition combined with an order-swapping property in time-ordered correlation functions
\begin{equation}
\langle \op_2(t_2) \op_1(t_1)\rangle = \langle e^{-\beta H}\op_1(t_1)e^{\beta H} \op_2(t_2)\rangle.
\label{kmscondition}
\end{equation}
It is in particular this order-reversing property that is quite subtle in its manifestation in our context. Therefore we will dissect its emergence in some detail in the present section. The correlators of interest in the Lorentzian regime are the HHL...L correlators of the form
\begin{equation}
\langle H|\op_n(t_n)...\op_1(t_1)|H\rangle,
\end{equation}
where all the light operators $\op_i$ are presumed to be time-ordered. Writing this correlator in Heisenberg form we find
\begin{equation}
\langle \op_H(0) \op_n(t_n)...\op_1(t_1)\op_H(0)\rangle = \langle \op_H(0) \op_n e^{-iH(t_n-t_{n-1})}\op_{n-1}...e^{-iH(t_2-t_1)}\op_1 \op_H(0)\rangle, 
\label{analyticcontinuation}
\end{equation}
where we have used the fact that the Hamiltonian annihilates primary states $H\op_H(0)|0\rangle = 0$. If we now introduce the parameters 
\begin{equation}
\tau_k = t_k-t_{k-1},    
\label{timedifferences}
\end{equation}
we can see that analytically extending $\tau_k$ to the lower half-plane yields a continuous expression for \eqref{analyticcontinuation}. By assumption, for real Lorentzian times, \eqref{analyticcontinuation} is nothing but a Wightman function and is therefore analytic in real values of $\tau_k$. Hence by the edge-of-the-wedge theorem \eqref{analyticcontinuation} extends to a holomorphic function of $\tau_k$ on \textit{the lower half-plane}.

In the next step, we give all the Lorentzian times small imaginary parts such that all the time-differences $\tau_k$ take value in the lower half-plane and subsequently set all the real parts equal to zero. This method, which was applied in \cite{Roberts:2014ifa} and subsequently utilized in \cite{Banerjee:2018tut}, ensures that once we analytically continue back to real Lorentzian times we evaluate the correlator with the desired operator ordering. In the resulting Euclidean correlator, all operators commute, the ordering of the original correlators is now encoded within the small imaginary parts of the light operators.

Finally, we take the Euclidean correlator from the cylinder to the radial plane 
\begin{equation}
\langle H|\op_1(z_1)...\op_n(z_n)|H\rangle,
\end{equation}
where $z_k= e^{i(t_k+\phi_k)}$. The assumption that all $\tau_k$ as definded in \eqref{timedifferences} take values in the lower half-plane translates to the following inequalities on the radial plane
\begin{equation}
|z_n|<|z_{n-1}|<...<|z_1|.
\label{operatorordering}
\end{equation}

To summarize all the steps we have taken into bullet points, the process was the following:
\begin{itemize}
    \item Set all Lorentzian real times equal to zero.
    \item Give each operator a Euclidean complex time such that the edge-of-the-wedge theorem ensures we obtain the correct time-ordering once we continue back to Lorentzian times in the final step.
    \item Evaluate the resulting Euclidean correlators.
    \item Continue the the real parts of the complex times back to their original values, set the imaginary parts to zero.
\end{itemize}
Now that we understand the effect of the Euclidean times on the overall operator ordering we have the means to understand the impact of shifting the Euclidean time label of one of the operators by $t_i\rightarrow t_i + i\beta$, or equivalently on the radial plane $z_i \rightarrow e^{\beta}z_i$. 

Crucially, the step where the time-ordering is determined by the imaginary values of the times takes place before the evaluation of the resulting Euclidean correlator. From the arguments of the previous section we establish that this correlation function in the high-temperature regime is dominated by a single conformal block which is a single-valued function on a thin cylinder. For this reason the values of the correlation function are left invariant by the KMS-condition, but, the labeling that determines the time-ordering lives on the universal cover of the cylinder. As a consequence, the value of the correlation function is left invariant by a full rotation of a single coordinate label of a light operator, but the ordering \eqref{operatorordering} is altered. Hence, the Lorentzian correlation function is subject to an operator reordering that implements the KMS-condition \eqref{kmscondition}.

\section{The Heavy-Light correlator as a GNS-linear form}
\label{sec:GNS}
Historically one of the powers of the GNS-construction has been to bring density matrices into the language of conventional quantum states \cite{Haag}. Within the AdS/CFT literature, the same has been applied for systems with a large separation of scales in their Hilbert space (e.g. HHL..L correlators at large $c$, large-$N$ expansion in SYM). 

So far we have constructed Lorentzian correlation functions using a two-step process. First, we compute an appropriate Euclidean correlator after setting the Lorentzian times equal to zero and next, we continue the Lorentzian times back to their original values. This two-step procedure can be realized in terms of a GNS-construction. For an introduction to the GNS-construction, we refer the reader to appendix \ref{sec:appendix-GNS}. We construct a GNS-linear form $\omega$ over the set of light operators, the latter denoted by $\mathcal{B}_L$. We construct this linear form in such a way that it computes the appropriate Euclidean correlator
\begin{equation}
\omega(\op(x)) = \langle H|\op(x)|H\rangle, \;\;\;\;\; \op(x)\in \mathcal{B}_L, \,\, x\in S^1,
\label{ourGNSform}
\end{equation}
hence we take $|H\rangle$ to be a heavy primary state in the large $c$ regime. Since we want to consider time-evolution from the Heisenberg picture, \textit{we restrict $x$ to lie on the unit circle.}
We apply this linear form to construct a GNS-Hilbert space out of the light operator algebra. We will utilize the GNS-linear form to construct an auxiliary Hilbert space, the GNS-Hilbert space. We will use the the linear form \eqref{ourGNSform} to define an inner product on the space $\mathcal{B}_L$:
\begin{equation}
\langle\langle \op_1(x_1)|\op_2(x_2)\rangle\rangle \equiv \omega(\op_1^{\dagger}(x_1)\op_2(x_2)) = \langle H|\op_1^{\dagger}(x_1)\op_2(x_2)|H\rangle.
\end{equation}
With this set-up, we are now in a position to apply the GNS-construction and associate a vector in a GNS-Hilbert space, to any operator in $\mathcal{B}_L$
\begin{equation}
O_i \longleftrightarrow |O_i\rangle\rangle.    
\end{equation}
We can immediately construct the GNS-vacuum state $|K_0\rangle\rangle$ which is automatically normalized
\begin{equation}
\langle\langle K_0|K_0\rangle\rangle = \langle H| \mathbb{I}^{\dagger} \mathbb{I}|H\rangle = \langle H|H\rangle = 1.
\end{equation}
We should of course also ensure that the scalar product is positive definite. In our case, this is guaranteed by construction since all spatial coordinates are restricted to the unit circle in the radial plane. More explicitly, we have
\begin{equation}
\langle\langle \op(x) | \op(x)\rangle\rangle = \langle H|\op^{\dagger}(1/\bar{x})\op(x)|H\rangle = |\op(x)|H\rangle|^2 \geq 0 
\end{equation}
Note that $|H\rangle$ is not annihilated by any of the light operators, as such the generated Hilbert space has trivial Gelfand-ideal, just like in the case of the KMS-state discussed in appendix \ref{sec:appendix-GNS}.

The distinctions and similarities between the auxiliary GNS-Hilbert space and the physical CFT Hilbert space are summarized in table \ref{GNS-CFTcomp}.

\begin{table}
    \centering
    \begin{tabular}{c|c}
    \hline
    CFT Hilbert space &  GNS-Hilbert space\\
    \hline
    Vectors: $|H\rangle$     & Vectors: $|O_i\rangle\rangle$\\
    Operators: $\mathcal{O}_i$ & Operators: $\pi(\mathcal{O}_i)$\\
    Vacuum: $|0\rangle$ & Vacuum: $|K_0\rangle\rangle$\\
    \hline
    \end{tabular}
    \caption{We will apply a double-bracket notation to distinguish between the auxiliary GNS-Hilbert space and the physical CFT Hilbert space.}
    \label{GNS-CFTcomp}
\end{table}

\subsection{representations of $\mathcal{B}_L$ over the GNS-Hilbert space}
We can now define two canonical representations $\pi(\mathcal{B})$ and $\pi_r(\mathcal{B})$ of $\mathcal{B}_L$ over the GNS-Hilbert space in the conventional manner
\begin{equation}
\pi(A)|B\rangle\rangle = |AB\rangle\rangle, \;\;\;\;\; \pi_r(A)|B\rangle\rangle = |BA^{\dagger}\rangle\rangle.
\end{equation}
It is easily verified that for all $A,B\in\mathcal{B}$
\begin{equation}
[\pi(A(x_1)),\pi_r(B(x_2))] = 0.
\end{equation}
The power of this construction is that this enables us to write the (non-vacuum) expectation value of the light operators on the heavy state
as a time-ordered vacuum expectation value of the canonical representation,
\begin{equation}
\langle H|\op(x_1)...\op(x_n)|H\rangle = \langle\langle \pi(\op(x_1))...\pi(\op(x_n))\rangle\rangle.
\label{representativecorrelator}
\end{equation}
This is also the reason for the introduction of the double-bracket notation $\langle\langle\dots\rangle\rangle$, so we can distinguish between the Euclidean correlation function of the CFT on the radial plane and the vacuum expectation values of representations of the light operators on the auxiliary GNS Hilbert space which computes the expectation values of the light operators on the heavy eigenstate on the Lorentzian cylinder.

The reason why this works is a little subtle but follows from the discussion on operator ordering of section 4.3, note that due to the definition of the GNS-linear form, that as long as the right-hand side of \eqref{representativecorrelator} is time-ordered, the right-hand side will automatically provide the Euclidean correlation function evaluated on the correct branch. 

The main point to drive home is that the role of the GNS construction here is effectively the same as that of the uniformization method exploited in earlier sections, namely, to reduce the HHL...L correlator to a correlator over (a representation of) the light operators. In other words, the GNS construction can be thought of as an algebraic manifestation of the uniformization map.

\subsection{GNS-construction in the $\alpha\gg h$ regime}
As we elaborated before, in the regime where $\alpha \gg h$ the HHL...L correlator is dominated by the identity block contribution. In this OPE limit, the uniformized correlator and the GNS-vacuum expectation value coincide
\begin{equation}
\langle\langle \pi(\op_1(x_1))...\pi(\op_n(x_n))\rangle\rangle \overset{\alpha\gg h}{=} \left(\frac{\partial w}{\partial z_1}\right)^{h_1}...\left(\frac{\partial w}{\partial z_n}\right)^{h_n}\langle \op(w(z_1)...\op(w(z_n))\rangle.
\end{equation}
As discussed in section \ref{sec:emergentKMS}, the right-hand side in this regime satisfies the KMS condition and is also time-translation invariant, hence, as a direct consequence, so does the left-hand side. We can construct a formal operator $\mathcal{K}$ on the GNS-Hilbert space, namely, the modular Hamiltonian, which generates translations in the Euclidean time-direction.

The modular Hamiltonian $\mathcal{K}$ has two properties, it annihilates the GNS-vacuum
\begin{equation}
\mathcal{K}|K_0\rangle\rangle =0,
\end{equation}
and it generators a one-dimensional automorphism group on the representatives of the light operators
\begin{equation}
\pi(\op)(\tau) = e^{\tau \mathcal{K}}\op e^{-\tau \mathcal{K}}.
\label{automorphism}
\end{equation}
In particular, this automorphism group satisfies the KMS condition
\begin{equation}
\langle\langle \pi(\op_2) \pi(\op_1)\rangle\rangle = \langle\langle e^{-\beta \mathcal{K}}\pi(\op_1)e^{\beta \mathcal{K}} \pi(\op_2)\rangle\rangle. 
\end{equation}
As a direct consequence, the modular Hamiltonian $\mathcal{K}$ satisfies an additional non-trivial property due to the uniqueness of automorphisms theorem \cite{Stratila1,Stratila2}. This theorem is reviewed in appendix \ref{sec:appendix-automorphism} and states that the operator $e^{-\beta \mathcal{K}}$, where $\beta=2\pi/\alpha$, has to coincide with the modular operator $\Delta$ which occurs in the Tomita-Takesaki operator $S$, where 
the Tomita-Takesaki operator $S$ has the property
\begin{equation}
|SOK_0\rangle\rangle = |O^{\dagger}K_0\rangle\rangle,
\label{TT}
\end{equation}
and the polar decomposition
\begin{equation}
S= J\Delta^{1/2}.
\end{equation}
A heuristic proof of this theorem is sketched in appendix \ref{sec:appendix-automorphism}. 
In turn, the operator $J$ is an anti-unitary operator\footnote{Anti-unitary implies that $\langle\langle x J|Jy\rangle\rangle = \langle\langle y|x\rangle\rangle$.} that satisfies the properties
\begin{align}
& |J \op K_0\rangle\rangle = |K_0 \op^{\dagger}\rangle\rangle, \label{antiunitary}\\
& J = J^{-1}.
\end{align}
From this, we derive a useful identity
\begin{equation}
e^{-\beta \mathcal{K}/2} = \Delta^{1/2} = JS.
\end{equation}
Combining \eqref{TT} and \eqref{antiunitary} yields
\begin{align}
|e^{-\beta \mathcal{K}/2} \op K_0\rangle\rangle = |J S \op K_0\rangle\rangle = 
 |J \op^{\dagger} K_0\rangle\rangle  = |K_0 \op\rangle\rangle.
\end{align}
We can pull this linear transformation of the GNS Hilbert algebra out of the brackets into an automorphism  that acts on the representative of the operator 
\begin{equation}
e^{-\beta \mathcal{K}/2} \pi(\op) e^{\beta \mathcal{K}/2}|K_0\rangle\rangle = \pi_r(\op^{\dagger})|K_0\rangle\rangle, 
\label{mirrormap}
\end{equation}
in the process of doing so we have reinterpreted this operator-switching property in terms of the commutant representation. It is worth noting that the automorphism $e^{-\beta \mathcal{K}/2}$ has the special property of mapping an operator to an operator in the commutant algebra. We will exploit this property in the next section.

\subsection{The non-equilibrium state and the GNS-construction}
We finally have the platform to state the main algebraic claim regarding the atypical non-equilibrium state \eqref{nonequilibrium} which is the focus of the present work. The modular operator $\exp(-\beta \mathcal{K}/2)$ has the property \eqref{mirrormap} of mapping an operator to the commutant algebra. We can relate $\mathcal{K}$ to the system Hamiltonian through
\begin{equation}
\mathcal{K} = \pi({\cal H}) - \pi_r({\cal H}).
\label{modhamiltonian}
\end{equation}
Since the identity operator commutes with $K_0$, applying \eqref{mirrormap} to a unitary operator $U\in\mathcal{B}$ gives
\begin{equation}
e^{-\beta \mathcal{K}/2} \pi(U) e^{\beta \mathcal{K}/2}|K_0\rangle\rangle  = \pi_r(U^{\dagger})|K_0\rangle\rangle = \pi_r(U)^{\dagger}|K_0\rangle\rangle.
\label{operatortocommutant}
\end{equation}  
Since the commutant representation of the Hamiltonian in \eqref{modhamiltonian} by definition commutes with $\pi(U)$, we can reduce the expression above to the more familiar form\footnote{Strictly speaking the operator $\pi({\cal H})$ does not have a well-defined large-$c$ limit.} 
\begin{equation}
e^{-\beta \mathcal{K}/2} \pi(U) e^{\beta \mathcal{K}/2}|K_0\rangle\rangle = e^{-\beta \pi({\cal H})/2} \pi(U) e^{\beta \pi({\cal H})/2}|K_0\rangle\rangle = \pi_r(U)^{\dagger}|K_0\rangle\rangle.
\end{equation}
Furthermore, one can easily show that
\begin{equation}
\pi_r(U)\pi_r(U)^\dagger |A\rangle = \pi_r(U)\pi_r(U^\dagger) |A\rangle = |AU^{\dagger} U\rangle = |A\rangle,  
\end{equation}
from which one can argue that $\pi_r(U)$ inherits the unitarity property of $U$.

This brings us to our final conclusion. Computing the light operator expectation values over the state $\exp(-\beta {\cal H}/2)\pi(U)\exp(\beta {\cal H}/2) |K_0\rangle$ yields
\begin{align}
& \langle\langle K_0|  e^{\beta {\cal H}/2}\pi(U)^{\dagger}e^{-\beta {\cal H}/2} \pi(\op(x_1))...\pi(\op(x_n))e^{-\beta {\cal H}/2}\pi(U)e^{\beta {\cal H}/2}|K_0\rangle\rangle \nonumber \\
& =\langle\langle K_0| \pi_r (U) \pi(\op(x_1))...\pi(\op(x_n)) \pi_r(U^{\dagger}) |K_0\rangle\rangle \nonumber \\
& =\langle\langle K_0| \pi_r(U)\pi_r(U)^{\dagger} \pi(\op(x_1))...\pi(\op(x_n))|K_0\rangle\rangle \nonumber \\
& =\langle\langle K_0| \pi(\op(x_1))...\pi(\op(x_n))|K_0\rangle\rangle.
\end{align} 
Hence, we are able to explicitly establish that these non-equilibrium states are actually indistinguishable from equilibrium states from the perspective of the light operators that make up the small algebra. This, indeed, is the defining property of these atypical non-equilibrium states\cite{Papadodimas:2017qit}.

\section{Discussion}
\label{sec:conclude}
Motivated by the behind-the-horizon excitation states of \cite{Papadodimas:2017qit}, we studied the properties of the expectation values of light CFT operators on heavy non-equilibrium states \eqref{nonequilibrium} and some general properties of expectation values on heavy equilibrium states. This was done in the regime of large $c$ and at large temperature $\alpha \propto\sqrt{H/c}\gg h$. We obtained some insights in a two-fold manner, first by performing some first-principle computations of conformal blocks in the large $c$, large $\alpha$ regime. Secondly, by implementing these computations at the GNS-level and borrowing the technology of Tomita-Takesaki theory we were able to identify the dressing of the unitary operators in \eqref{nonequilibrium} as a procedure that maps its operator representative to the commutant algebra, which essentially maps $U$ to its mirror operator \cite{Papadodimas:2013jku}. We close off with a discussion of a set of questions/thoughts induced by this conclusion.   

\subsection{The GNS-construction as coarse-graining}
A common, interesting question regards the emergence of semi-classical gravity from conformal field theory. It is well-understood that the regime where we can consider Fock spaces of non-backreacting particle excitations on top of a curved background corresponds to CFT with a large number of degrees of freedom ('t Hooft limit in 4d $\mathcal{N}=4$ SYM, or large-$c$ in 2d CFT). In this case the subspace of the boundary CFT Hilbert space corresponding to this Fock space is given by the space generated by acting with an algebra of `simple' operators on some complicated CFT state corresponding to the bulk boundary geometry \cite{Papadodimas:2012aq}. Familiar cases include $\mathcal{N}=4$ SYM, where these would be the single-trace operators, and 2d CFT case relevant to this article where we studied the `light' operators.

Reducing the complicated, dense, full Hilbert space of the CFT to this effective subspace in a natural manner induces a form of coarse-graining to a set of observables that are accessible to a fictitious experimentalist with minimal resolution power. The GNS-construction, where we consider representatives of the light operators acting on an auxiliary Hilbert space, provides for us a useful formal language in which to express this coarse-graining. For one, it is clear that the GNS-Hilbert space we constructed is exactly the Fock space-like Hilbert subspace we described above. 

We have argued for the application of the GNS-construction to take over the role of the two-step procedure that we utilized in the past \cite{Vos:2018vwv,Banerjee:2018tut}. In our past approach we compute the conformal blocks at the Euclidean level and subsequently analytically continue to Lorentzian times, this automatically raises the question of time-ordering ambiguities, which we dealt with at the Euclidean level by giving all operators small imaginary times that encode their ordering.

In the GNS-approach utilized in this construction the operators that generate the heavy states are still given Euclidean times corresponding to their place in the future and past respectively, but we chose to give all light operators the same Euclidean time. The price we pay for this inability to resolve the ordering ambiguity is that at the GNS-level, where we compute vacuum expectation values of the representatives of the operators, we have to demand that all the representatives are time-ordered. Hence we have essentially shifted the burden of resolving the time-ordering ambiguity from the first to the second step. This is essentially a weakness, we gave up on the small imaginary parts in the light-operator spectrum and therefore we had to reintroduce the concept of time-ordering at the GNS-space level. But it does come with a side-effect, since switching the operator orderings implicitly changes the time-ordering, we find that our algebra has no non-trivial center, as such it forms a von Neumann factor \cite{Haag} which raises some potentially interesting questions.

\subsection{Classification of the light-operator von Neumann factor.}
Having established that our light operator von Neumann algebra forms a factor it is an interesting question to try and identify what kind of factor our algebra is in terms of their classification. One reason being the key role played by the commutant algebra in our analysis. If the algebra were to be type I then there is a possibility that our GNS-linear form is `pure', in this context meaning that the algebra acts irreducibly on the GNS-Hilbert space which signals that the commutant algebra is trivial\footnote{note the distinction with our earlier use of the term pure, where we simply meant that our primary state $|H\rangle$ is a pure state of the full CFT Hilbert space}. If our algebra is either type II or III it is guaranteed to act reducibly on the GNS-Hilbert space which allows for a non-trivial commutant algebra, the physical interpretation is that for an algebra of type II or III any state looks like a mixed state. This was used in the mirror map construction in \cite{Papadodimas:2013jku} where it was explicitly shown that the Hilbert space is sufficiently large to accommodate the commutant algebra, i.e. the mirror operators.

While this is a difficult question to answer from first principles (the well-known recent method being to identify a family of half-sided modular inclusions \cite{Leutheusser:2021frk}), we can make some comments. All automorphisms of type I algebras are inner automorphisms \cite{Witten:2021unn}. If the generator $\mathcal{K}$ were to generate an inner automorphism this would imply that $\mathcal{K} \in \pi(\mathcal{B}_L)$ which would imply that $[\mathcal{K},\pi_r(O_i)]=0$, for all $O_i\in\mathcal{B}_L$. Applying this to expression \eqref{operatortocommutant} and reminding ourselves that $\mathcal{K}$ annihilates the GNS-vacuum would give us
\begin{equation}
\pi(U)|K_0\rangle\rangle = \pi_r(U)^{\dagger}|K_0\rangle\rangle,    
\end{equation}
which would in turn imply that \textit{all} quantum quenches would be unobservable to the light observable algebra. The only way to avoid this outcome is to conclude that \eqref{automorphism} is an \textit{outer} automorphism. Given that the GNS-linear form we used generates a faithful representation $\pi(\mathcal{B}_L)$ we conclude that $\mathcal{B}_L$ has to be either type II or type III.

\subsection{The Cardy regime, the Hagedorn transition and OPE limits}
We observed thermal behavior at the conformal block level in the ultra-high energy regime, where we found the temperature to be very large compared to the scaling dimensions of the light operators. This puts this energy range firmly in the Cardy regime, which begs the natural question: is this not just a consequence of modular invariance? 

To some extent it would seem so, we found that in the limit $\alpha\gg h$ the vacuum conformal block of expectation value $\langle H|\op...\op|H\rangle$ is given by a conformal block of the vacuum expectation value $\langle \op...\op\rangle$ on a geometry which is a very thin cylinder with a compactified time direction. By modular invariance this is related by a discrete transformation to the vacuum expectation value. There are two major factors that differentiate our analysis from results that could be obtained through a modular invariance. 

Firstly, from the ground up our analysis is based on expectation values of light operators on a heavy eigenstate of the Hamiltonian, we did not assume that this state is a typical state of a thermal ensemble, we derived thermal properties of these expectation values. 

Secondly, our limit is taken from a different direction. Instead of taking the low-energy limit and applying a discrete transformation we instead take the high-energy regime ($H\sim c$) and subsequently take the ultra high-energy limit ($\sqrt{H/c} \gg h$). This gives us some potential perspective into a possible Hagedorn transition in large-$c$ CFT. Our analysis is based on universal features of the Virasoro algebra, particularly the Virasoro Ward identity. 

As stated before, on its own, this teaches us nothing new as this is simply in compliance with the Cardy formula. What is interesting is that the limit to the Cardy regime is essentially smooth with no indication of a phase transition, which, if there is a low-energy generalized free sector in the CFT, would be in contradiction with the predictions of \cite{El-Showk:2011yvt} . There is another transition that is sharp though, at $H \geq c/24$ the combination $\alpha= \sqrt{24 H/c-1}$ becomes real-valued. In models that are holographically dual to AdS$_3$ this corresponds to a bulk object with an ADM mass that reaches the minimal BTZ threshold \cite{Aharony:1999ti}. This suggests that these models possess an extended Cardy regime as described in \cite{El-Showk:2011yvt}.

This leads to a potentially more natural holographic interpretation. We found that in the Cardy regime $\alpha\gg h$ the lowest-lying conformal block takes on thermal properties, this is on its own not a statement on the thermal properties of the full conformal block expansion of the correlator. What we do find is that the validity of our coordinate range in the thin-cylinder regime is roughly $|\tilde{\phi}|,|\tilde{\tau}|< n|h/\alpha|$, where $n$ is some $\mathcal{O}(1)$ constant. Hence if we assume a gap between the lowest-lying non-trivial primary state and the vacuum (as is required in order to have a semi-classical gravitational dual \cite{El-Showk:2011yvt}) that the lowest-lying conformal block gets projected out of the full correlator as a direct consequence of the implicit OPE limit, hence we suggest that the full correlator might be thermal all the way down to at the energy scale $H>c/24$ \cite{Fitzpatrick:2015zha,Banerjee:2018tut,Vos:2018vwv}, but that it is in the Cardy regime that the single conformal block dominates the correlator and as a consequence has to inherit the thermal properties of the full correlator.

\subsection{The algebraic approximation} The emergent notion of KMS-condition in the limit $\alpha \gg h$ as discussed in section \ref{sec:emergentKMS} or equivalently, that of the commutant algebra in the same limit, as advocated in terms of the GNS-linear form in section \ref{sec:GNS}, has an interesting interpretation in terms of the underlying von Neumann algebra. Our starting point was a generic state vector in the full quantum theory and therefore the associated algebra of operators can be expected to belong to type I. Starting with this, we perform a limiting procedure of setting the central charge large such that it creates a hierarchy between operators, based on the parametric dependence of their respective scaling dimensions with the central charge. Following the uniformization technique, such a hierarchy leads naturally to the emergent KMS-condition interpreted as an effective thermality. In terms of the GNS-linear form, the same limit yields a non-trivial commutant algebra, once again signalling an effective thermality. It is interesting to note that the KMS-condition can indeed be interpreted in terms of a lack of a tracial state \cite{Witten:2023qsv}, thus interpreting this emergent thermality as an effective algebraic transition between a type I algebra and a type III algebra. The latter is only realized in a limiting sense. Such type III approximations of type I algebras have recently been discussed in the context of understanding emergent thermality in brick wall models in \cite{Banerjee:2024dpl, Burman:2023kko, Soni:2023fke}. We expect that a more precise understanding of this algebraic approximation might play a pivotal role in understanding the pure-state description of black holes and the so-called factorization puzzle \cite{Jafferis:2019wkd}. We hope to come back to this in a near-future work.

\section*{Acknowledgements}
We would like to thank Kyriakos Papadodimas for the initial collaboration and several useful discussions. We would also like to thank him further, for his extremely insightful comments on an earlier version of this manuscript. In addition, we thank Chethan Krishnan, Joris Raeymaekers and Paolo Rossi for helpful discussions. The work of GV was supported by the Grant Agency of the Czech Republic under the grant EXPRO 20-25775X.

\appendix
\section{The GNS-construction}
\label{sec:appendix-GNS}
Given a linear form $\omega$ over a closed $*$-algebra $\mathcal{A}$, we can construct a Hilbert space $\mathcal{H}$ out of $\mathcal{A}$, in addition we can construct a representation $\pi_{\omega}(\mathcal{A})$ of $\mathcal{A}$ of linear operators acting on $\mathcal{H}$. This procedure somewhat analogous to how we construct the adjoint representation out of the fundamental representation in representation theory. The process of building a representation out of $\mathcal{A}$ in this manner is called the GNS-construction and it plays a central role in section 5 hence we will briefly review its procedure based on chapter III of \cite{Haag}.

Assume the existence of a positive linear form $\omega$ on $\mathcal{A}$, positivity in this context implies that 
\begin{equation}
\omega(A^{*}A) \geq 0.
\end{equation}
As a concrete example of the form $\omega$ we can think of a density operator combined with the trace operation, i.e. if $\rho$ is density operator than the linear form $\rho(A)$ defined as
\begin{equation}
\rho(A)=\text{Tr}(\rho A),
\end{equation}
satisfies the necessary criteria for the form $\omega$.

We can use $\omega$ to construct a scalar product on $\mathcal{A}$, if $A,B \in \mathcal{A}$ then
\begin{equation}
\langle \langle A|B\rangle\rangle \equiv \omega( A^{*}B).
\end{equation} 
As this is a proper scalar product in the formal sense it satisfies the usual inequality
\begin{equation}
|\langle\langle A|B\rangle\rangle|^2 \leq \langle\langle A|A\rangle\rangle \langle\langle B|B\rangle\rangle.
\label{angleinequality}
\end{equation}
The resulting linear space is not quite a Hilbert space yet as the null-vector is not yet unique. To fix this consider the subalgebra $\mathcal{J} \subset \mathcal{A}$, which is defined as
\begin{equation}
X\in \mathcal{J} \;\;\; \Rightarrow \;\;\; \langle\langle X|X\rangle \rangle = 0
\end{equation}
The inequality \eqref{angleinequality} implies that $\mathcal{J}$ is a left-invariant subideal of $\mathcal{A}$. $\mathcal{J}$ forms the so-called Gelfand ideal of $\omega$. As a result the quotient $\mathcal{A}/\mathcal{J}$ is a Hilbert space\footnote{Technically its completion under the norm topology induced by the scalar product forms the Hilbert space.}, its elements are the equivalence classes $|A\rangle\rangle$, i.e. if $A\in|A\rangle\rangle$ then $A+X \in |A\rangle\rangle$ for any $X\in \mathcal{J}$.

This is where the construction comes in, for any element $A\in\mathcal{A}$ we can define a linear operator $\pi_{\omega}(A)$ acting on vectors $|B\rangle\rangle\in \mathcal{A}/\mathcal{J}$ trough
\begin{equation}
\pi_{\omega}(A)|B\rangle\rangle \equiv |AB\rangle\rangle.
\end{equation}
The big advantage of this construction is that it can be utilized to represent expectation values of density matrices in a manner reminiscent of pure-state quantum mechanics. Take the state $\Omega \in \mathcal{A}/\mathcal{J}$ to be the class $[\mathbb{I}]$, i.e. the equivalence class that contains the identity operator, then
\begin{equation}
\omega(A) = \langle\langle \Omega|\pi_{\omega}(A)|\Omega\rangle\rangle.
\end{equation}
It is said that the state $|\Omega\rangle$ represents the `state' $\omega$. 

\subsection{Left- and right-GNS construction}
The construction described above is not the full story, one can construct another representation $\pi_{r}(A)$ whose weak closure is exactly the commutant algebra of $\pi(A)$. The right-representation $\pi_r(A)$ is defined through
\begin{equation}
\pi_r(A) |x\rangle\rangle = |xA^{*}\rangle\rangle,
\end{equation}
i.e. we right multiply with the conjugate\footnote{Hence the original representation we defined in the previous section should be called $\pi_l(A)$, \cite{Haag} argues that the subscript can be dropped since for the KMS state we will be interested in in the next section $\pi(\mathcal{R})$ is isomorphic to $\mathcal{R}$. I will maintain consistency with the notation of \cite{Haag}}. This has the immediate property that any operator $\pi_r(B)$ commutes with any operator $\pi(A)$:
\begin{equation}
\pi_r(B)\pi(A)|x\rangle\rangle = \pi_r(B)|Ax\rangle\rangle = |AxB^{*}\rangle\rangle = \pi(A)|xB^{*}\rangle\rangle = \pi(A)\pi_r(B)|x\rangle\rangle.
\end{equation}
In fact, as mentioned above that (the weak closure of) the representation $\pi_r(\mathcal{R})$ is the commutant of $\pi(\mathcal{R})$. 

We can construct an anti-unitary\footnote{For completenes, anti-unitary implies that $\langle\langle Jx|Jy\rangle\rangle = \langle\langle y|x\rangle\rangle$.} operator $J$ that has the property that it maps $\pi(\mathcal{R})$ to $\pi_r(\mathcal{R})$. In order to do that define $J$ such that
\begin{equation}
|J x\rangle\rangle = |x^{*}\rangle\rangle.
\label{antiunitaryJ}
\end{equation}
We can now realize the operator $J$ at the level of representations, strictly speaking we would denote this operator as $\pi(J)$, but we will keep the notation light and just refer to it as $J$ as well, we hope this does not lead to confusion. In this case, it is now easily shown that
\begin{equation}
J\pi(A)J = \pi_r(A),  \;\;\;\; J^2 = \mathbb{I},
\end{equation}
the proof is straightforward
\begin{equation}
J\pi(A)J|x\rangle\rangle = J\pi(A)|Jx\rangle\rangle  = J\pi(A)|x^{*}\rangle\rangle = J|Ax^{*}\rangle\rangle = |xA^{*}\rangle\rangle = \pi_r(A)|x\rangle\rangle.
\end{equation} 
This operator $J$ takes part in the polar decomposition of the Tomita-Takesaki operator $S$ discussed in section \ref{sec:GNS}.

\subsection{The KMS-state}
The KMS state $\omega_{\beta}(A)$ is defined as the linear form on the space of operators that produces thermal expectation values with inverse temperature $\beta$
\begin{equation}
\omega_{\beta}(A)=Z^{-1}\text{Tr}\left(e^{-\beta {\cal H}} A\right).
\end{equation}
If we define time-evolution of an operator through
\begin{equation}
a_t A \equiv e^{i{\cal H}t}Ae^{-i{\cal H}t}
\label{timeevo}
\end{equation}
then it is trivial to show that
\begin{equation}
\omega_{\beta}(a_t A) = \omega_{\beta}(A).
\end{equation}
This demonstrates the usual time-translation invariance of operators in thermal equilibrium. What is more interesting is when one operator is evolved with respect to another operator, in which case we find 
\begin{align}
&\omega_{\beta} ((a_t A) B) = Z^{-1} \text{Tr}\left(e^{-\beta {\cal H}} e^{i{\cal H}t}Ae^{-i{\cal H}t} B\right) \nonumber \\
&= Z^{-1}\text{Tr}\left(Be^{i{\cal H}(t+i\beta)} A e^{-i{\cal H}t}e^{\beta {\cal H}}e^{-\beta {\cal H}}\right) \nonumber \\
&= Z^{-1}\text{Tr}\left(e^{-\beta {\cal H}} B e^{i{\cal H}(t+i\beta)} A e^{-i{\cal H}(t+i\beta)}\right) \nonumber \\
&= \omega_{\beta}(Ba_{t+i\beta}A).
\end{align}
This is the KMS-condition, it implies a variety of important analyticity properties of the thermal state on a strip, but in the main body of the text we will mostly be concerned with its application to Tomita-Takesaki theory.

\subsection{GNS-construction applied to the KMS-state}
We will now use the power of the GNS-construction on the KMS-state. First a note on the relevant operator algebra, we will want observables that have a well-defined limit in the thermodyncamic limit. This means we will consider the observables $A\in B(\mathcal{H}_F)$, i.e. the space of bounded operators acting on the multiparticle Fock space build on the thermal state. We will be specific on this point as it will lead to some subtleties as we will see and these subtleties are either exactly the same or analogous in the large-$c$ limit we will consider later on. 

First, we will split the density matrix with an operator $K_0$ such that
\begin{equation}
K_0 = e^{-\beta {\cal H}/2}.
\end{equation}
We can now define the scalar product 
\begin{equation}
\langle\langle K_0 | K_0\rangle\rangle = \text{Tr}(e^{-\beta {\cal H}}).
\end{equation}
With this definition, thermal expectation values $\omega(A)$ take form of expectation values over the GNS-state
\begin{equation}
\omega(A) \equiv \text{Tr}\left(e^{-\beta {\cal H}} A\right) = \langle\langle K_0 |A|K_0 \rangle\rangle.
\end{equation}
And we define the algebra representations
\begin{equation}
\pi(A)|K_0\rangle\rangle = |Ae^{-\beta {\cal H}/2}\rangle\rangle, \;\;\;\;\; \pi_{r}(A) |K_0\rangle\rangle = |e^{-\beta {\cal H}/2} A^{*}\rangle\rangle.
\end{equation}
One particular property of this particular linear form on the space $B(\mathcal{H}_F)$ is that no operators in the fock space have vanishing thermal expectation value, i.e.
\begin{equation}
\text{if} \; A\in B(\mathcal{H}_F), \;\;\;\;\; \text{then}\;\; \omega(A) \neq 0,
\end{equation}
as a result the GNS-vector space has an empty Gelfand ideal and is isomorphic to $B(\mathcal{H}_F)$.

Another salient point of thermal expectation values, they are time-translation invariant
\begin{equation}
\langle\langle K_0 | a_t A|K_0\rangle\rangle = \omega(a_t A) = \omega(A),
\end{equation}
with time-translations defined as in \eqref{timeevo}. As a result there exists a unitary automorphism U(t) on the formal state $|K_0\rangle$ that leaves the state invariant
\begin{equation}
U(t)|K_0\rangle\rangle = |K_0 \rangle\rangle.
\end{equation}
We can associate a Hermitian generator $\mathcal{K}$ to this one-parameter group of automorphisms in the usual way
\begin{equation}
U(t) = e^{-it \mathcal{K}}.
\end{equation}
Note that while $\mathcal{K}$ is the generator of time-translation invariance of the formal state $|K_0\rangle$ it is definitely \textit{not} the Hamiltonian ${\cal H}$. This is easily seen from the fact that by construction
\begin{equation}
\mathcal{K}|K_0\rangle\rangle =0,
\label{timetranslation}
\end{equation}
which is inconsistent with the Hamiltonian, which has the property
\begin{equation}
\langle\langle K_0|{\cal H}|K_0\rangle\rangle = \text{Tr}\left(e^{-\beta {\cal H}/2} H e^{-\beta {\cal H}/2}\right) = E.
\end{equation}
In fact, $E$ is an extrinsic quantity in the thermodynamic limit, as a result $\mathcal{H}|K_0\rangle$ is not even a normalizable state in the thermodynamic limit. Despite this shortcoming, we can construct the operator $\mathcal{K}$ out of the representations of the Hamiltonian. Take $\mathcal{K}$ to be the following linear combination
\begin{equation}
\mathcal{K} = \pi({\cal H}) - \pi_r ({\cal H}).
\end{equation}
It follows straightforwardly that this linear combination has the correct annihilation property \eqref{timetranslation}
\begin{align}
& \langle\langle A| \left(\pi({\cal H}) - \pi_r ({\cal H})\right)|K_0\rangle\rangle = \text{Tr}\left( e^{-\beta {\cal H}/2}A^{*} {\cal H} e^{-\beta {\cal H}/2}\right) - \text{Tr}\left( e^{-\beta {\cal H}/2}A^{*}  e^{-\beta {\cal H}/2} {\cal H}^{*}\right) \nonumber \\
& \text{Tr}\left( e^{-\beta {\cal H}/2}A^{*} \left[ {\cal H}, e^{-\beta {\cal H}/2}\right]\right) =0.
\end{align}
This linear combination is very familiar. It is exactly how we construct the full boundary Hamiltonian of the eternal black out of the modular Hamiltonians of the left and right wedge. Note especially, that because the representations $\pi(A)$ and $\pi_r(A)$ commute with one another it satisfies another familiar feature of the modular Hamiltonian
\begin{equation}
\frac{\partial}{\partial t} \pi(a_t A) |_{t=0} = i[\mathcal{K},\pi(A)] = i[\pi({\cal H}),\pi(A)].
\end{equation}  
Note that on its own $\pi({\cal H})$ is as ill-defined in the thermodynamic limit (or large-$N$) limit as $H$ itself. Whereas the subtracted version $\mathcal{K}$ is well-defined in the thermodynamic limit. This observation plays a key role in \cite{Leutheusser:2021frk}.

In terms of the representatives of the Hamiltonian we can write the formal evolution operator generated by $\mathcal{K}$ as
\begin{equation}
e^{-\beta \mathcal{K}/2} = \pi\left(e^{-\beta {\cal H}/2}\right)\pi_r\left(e^{\beta {\cal H}/2}\right).
\end{equation}
From this we establish the property
\begin{align}
& e^{-\beta \mathcal{K}/2} |A K_0\rangle\rangle = \pi\left(e^{-\beta {\cal H}/2}\right)\pi_r\left(e^{\beta {\cal H}/2}\right) |A e^{-\beta {\cal H}/2}\rangle\rangle \nonumber \\
& =|e^{-\beta {\cal H}/2} A e^{-\beta {\cal H}/2} e^{\beta {\cal H}/2}\rangle\rangle \nonumber\\
& =|K_0 A \rangle\rangle.
\end{align}
Taking the anti-unitary operator $J$ to be defined as in \eqref{antiunitaryJ}, we find 
\begin{equation}
e^{-\beta \mathcal{K}/2} |AK_0\rangle\rangle = J| A^{*}K_0\rangle\rangle,
\end{equation}
from which we immediately obtain
\begin{equation}
\pi(A^{*})|K_0\rangle\rangle = Je^{\beta \mathcal{K}/2} \pi(A)|K_0\rangle\rangle,
\end{equation}
Which is a direct manifestation of the anti-linear $S$-operator of the Tomita-Takesaki theorem in polar decomposition form.

\section{Automorphism uniqueness theorem}
\label{sec:appendix-automorphism}
In this appendix we provide a heuristic physicist's proof for the automorphism uniqueness theorem applied in section 5. For a rigorous proof of this theorem we will refer the reader to section 10.16 of \cite{Stratila1}.

We assume the existence of a one-parameter automorhpism group $p_t$ on the representatives of the algebra
\begin{equation}
p_t \pi(\op) = e^{it\mathcal{K}}\pi(\op)e^{-it\mathcal{K}},    
\end{equation}
with the property that it satisfies the KMS-condition
\begin{equation}
\langle\langle \pi(\op_1)p_{t-i\beta}\pi(\op_2)\rangle\rangle = \langle p_t\pi(\op_2)\pi(\op_1)\rangle\rangle. 
\end{equation}
From the presumed continuity of the one-parameter automorphism group we establish that there exists a continuous mapping $u_t$ on the GNS-Hilbert space such that
\begin{equation}
p_t\pi(\op) = \pi(u_tO). 
\end{equation}
If $S$ is the Tomita-Takesagi operator we find
\begin{align}
&\langle\langle \op_1|u_{t-i\beta}\op_2\rangle\rangle = \langle\langle \pi(S\op_1)p_{t-i\beta}\pi(\op_2)\rangle\rangle = \langle\langle p_{t}\pi(\op_2)\pi(S\op_1)\rangle\rangle \nonumber\\
&=\langle\langle Su_t \op_2|S\op_1\rangle\rangle = \langle\langle J\Delta^{1/2}u_t \op_2|J\Delta^{1/2}\op_1\rangle\rangle = \langle\langle \Delta^{1/2} \op_1|\Delta^{1/2}u_t\op_2\rangle\rangle \nonumber\\
& =\langle\langle  \op_1|\Delta u_t\op_2\rangle\rangle.
\label{KMSdelta}
\end{align}
We can represent the unitary evolution $u_t$ through means of a Hermitean operator which in anticipation we will call $\mathcal{K}$
\begin{equation}
|u_t \op\rangle\rangle = |e^{-i\mathcal{K}t}\op\rangle\rangle.
\end{equation}
From \eqref{KMSdelta} we now find
\begin{equation}
|u_{t-i\beta}\op\rangle\rangle = |e^{- \mathcal{K}\beta}e^{-i\mathcal{K}t}\op\rangle\rangle = |\Delta e^{-i\mathcal{K}t} \op\rangle\rangle. 
\end{equation}
From this we conclude
\begin{equation}
|\Delta^{1/2} \op\rangle\rangle= |e^{-\frac{\beta}{2} \mathcal{K}}\op\rangle\rangle
\end{equation}
We can pull this automorphism acting on the GNS-Hilbert space back to an automorphism acting on the algebra representation and conclude
\begin{equation}
e^{it\beta\mathcal{K}}\pi(\op)e^{-it\beta\mathcal{K}} = \Delta^{it}\pi(\op)\Delta^{-it}.
\end{equation}
Hence we find that the generator of any time-translation invariant one-parameter automorphism group that satisfies the KMS-condition is related by a simple formula to the modular operator in the polar decomposition of the Tomita-Tagesaki operator.

\bibliography{JHEPresubmissionBehindTheHorizon}
\bibliographystyle{JHEP}

\end{document}